\documentclass[aps,twocolumn,showpacs,superscriptaddress]{revtex4}
\usepackage{graphicx}

\begin{document}

\newcommand{\be}   {\begin{equation}}
\newcommand{\ee}   {\end{equation}}
\newcommand{\ba}   {\begin{eqnarray}}
\newcommand{\ea}   {\end{eqnarray}}

\title{On the classical-quantum correspondence for the scattering dwell time}

\author{Caio H. Lewenkopf}
\email{caio@uerj.br}
\homepage{http://www.dft.if.uerj.br/usuarios/caio}
\affiliation{ Instituto de F\'{\i}sica,
            Universidade do Estado do Rio de Janeiro,\\
            Rua S\~ao Francisco Xavier 524, 20550-900 Rio de Janeiro, Brazil }

\author{Ra\'ul O. Vallejos}
\email{vallejos@cbpf.br}
\homepage{http://www.cbpf.br/~vallejos}
\affiliation{ Centro Brasileiro de Pesquisas F\'{\i}sicas (CBPF), \\
              Rua Dr.~Xavier Sigaud 150, 22290-180 Rio de Janeiro, Brazil}

\date{\today}

\begin{abstract}
Using results from the theory of dynamical systems, we derive a general 
expression for the classical average scattering dwell time $\langle\tau\rangle$. 
Remarkably, $\langle\tau\rangle$ depends only on a ratio of phase space 
volumes.
We further show that, for a wide class of systems, the average classical 
dwell time is not in correspondence with the energy average of the quantum 
Wigner time delay.
\end{abstract}



\maketitle

\section{Introduction}

The study of the time a quantum collision process takes to occur 
is one of the most interesting chapters in scattering theory.
This problem turns out to be  subtle and fascinating due to the 
lack of a Hermitian operator to measure the time as a quantum 
observable. 
Hence, one inevitably has to rely on auxiliary constructions to 
quantify the time spent by a scattering process. 
To that end several ingenious strategies have been proposed over the 
past 50 years \cite{Carvalho02}. 
In a pioneering work, Eisenbud and Wigner \cite{Wigner55} proposed 
measuring the scattering delay time by recording the peak position 
of wave packets scattered in one-dimension. 
This simple construction, that just invokes the concept of group velocity, 
already captures the deep connection between the energy variations of the 
scattering phase shift and the delay time.
In 1960, Smith \cite{Smith60} put forward an alternative scheme,
applicable to stationary scattering processes, where the 
scattering dwell time is associated to the ratio between the 
probability of finding the particle inside the scattering region 
and the flux through its surface.
This approach has the advantage of eliminating the necessity of 
wave-packets and can be easily generalized to multi-channel 
scattering. 
As a result, the dwell-time $\tau_W(E)$ is expressed as
\be
\tau_W(E)= -\frac{i \hbar}{N} \sum_{a,b=1}^N {\sf S}_{ab}^\ast 
\frac{\partial {\sf S}_{ba}}{\partial E}  ~,
\label{eq:Wigner-time-delay}
\ee
where the scattering matrix ${\sf S}$, that encodes all accessible information 
about the scattering process, is taken at the energy $E$. The sums in
(\ref{eq:Wigner-time-delay}) run over all $N$ open asymptotic scattering 
channels. The time $\tau_W(E)$ is usually called Wigner time 
delay.

Is $\tau_W(E)$ in correspondence with the classical dwell time 
for general scattering systems?
To answer this question we approach the problem from the classical
side. 
We use the theory of dynamical systems to obtain a remarkably
simple and general expression for the classical dwell time,
revealing its geometric nature.
%
%
Comparing this result with the semiclassical limit for the energy 
averaged quantum dwell-time, we find that the quantum-classical 
correspondence does not hold in general. 

The present analysis does not contradict a previous study of 
ours \cite{Lewenkopf04}. 
There we followed a different path, applicable only to chaotic 
systems, and concluded that the classical-quantum correspondence 
for the dwell time holds. 
Here, approaching the problem in a way that is insensitive to the 
details of the dynamics,
we vastly expand \cite{Lewenkopf04} and show that the correspondence 
fails in the more general case of systems with mixed phase space.

The paper is organized as follows.
In Sec.~\ref{sec2} we derive the central result of this paper, namely,
a general expression for the classical average time delay in terms of 
the system phase space volume.
It is key to our analysis the formulation of the scattering process 
as the first return of a measure-preserving map, allowing us to benefit 
from well-known results of ergodic theory.
In Sec.~\ref{sec:quantum-time-delay} we discuss the semiclassical limit of
the energy-averaged Wigner time delay. 
We conclude presenting, in Sec.~\ref{sec:conclusions}, a comparison 
between the classical and quantum dwell time. We show that, in general, 
these two quantities do not coincide.

\section{The classical dwell time}
\label{sec2}

Poincar\'e sections are extremely useful tools for the analysis of 
phase space structures in bounded low-dimensional Hamiltonian systems:
These surfaces of section allow us to reduce the continuous time 
evolution of dynamical systems to discrete mappings, much simpler 
to work with.

Surfaces of section are essential for the proper definition of a
scattering problem. Consider the scattering of a particle by a
potential. The description of the scattering process requires two 
control surfaces for detecting the state of the particle before and 
after the scattering event.
The description of all possible scattering processes demands the 
control surfaces to be chosen so as to enclose the scatterer 
completely.
In this case we can consider just one surface for registering
both the states of incoming and outgoing particles.

Let us illustrate these concepts by discussing a generic scattering 
process in three-dimensions. We choose a spherical control
surface enclosing the region where the potential is non-negligible.
A point on the associated Poincar\'e surface $\Sigma$ has
coordinates $({\bf q},{\bf p}_\Vert)$, where $\bf q$ represents a
position on the sphere and ${\bf p}_\Vert$ the conjugate (angular)
momentum.
An incoming state is completely specified by giving its coordinates
on $\Sigma$ together with the condition that the momentum normal to
the sphere, ${\bf p}_\perp$, must point inwards
(the modulus of ${\bf p}_\perp$ is fixed by energy conservation).
The incoming state then evolves inside the scattering region, along a
trajectory given by Hamilton equations. It eventually intersects
$\Sigma$ again at the exit point $({\bf q}^\prime,{\bf p}_\Vert^\prime)$
and escapes.
Hence, any scattering process can be essentially viewed as the first 
return map of $\Sigma$ \cite{Ozorio00},
\begin{equation}
\label{poincare}
{\cal S}: \Sigma \mapsto \Sigma \; , \qquad
({\bf q},{\bf p}_\Vert)
\stackrel{\cal S}{\longmapsto}
({\bf q}^\prime,{\bf p}_\Vert^\prime) \; .
\end{equation}
As a consequence of the Poincar\'e-Cartan theorem, this map is
volume-preserving \cite{Ozorio88}.

The structure of the classical scattering problem has a clear 
quantum mechanical counterpart.
The quantum analogue of the classical Poincar\'e surface is the 
Hilbert space ${\cal H}_\Sigma$
associated to $\Sigma$. The quantum scattering ${\sf S}$-matrix is
a linear operator of ${\cal H}_\Sigma$, mapping incoming states
into outgoings ones. The Poincar\'e map (\ref{poincare}) is the
classical limit of ${\sf S}$. Conversely, the scattering matrix
${\sf S}$ can be thought of as the quantization of ${\cal S}$
\cite{Rouvinez95}. 
The unitarity of ${\sf S}$ is the quantum counterpart of the classical
volume conservation \cite{Miller74,Smilansky91}.

This parallel between classical and quantum scattering processes
serves to facilitate the determination of some quantum-classical
correspondences. For instance, and very useful for what follows, 
it becomes clear that the classical analogue
of an average over ``channels" (a complete basis set of 
${\cal H}_\Sigma$) is an average over $\Sigma$ weighted by its 
Liouville measure.

Let us now discuss in detail a very simple scattering system:
a two-dimensional billiard with an attached pipe.
The case of a smooth cavity with several (smooth) pipes
in two or three dimensions, or even, the scattering of
asymptotically free particles by a smooth potential,
are conceptually equivalent to the two-dimensional
billiard with a single pipe, and will be discussed later.

The physical process we analyze is:
A classical particle propagates along the pipe
and eventually arrives at the billiard, where it elastically
bounces $n$ times at the walls before escaping (see Figure~\ref{fig1}).
A Poincar\'e section $\Sigma$, transverse to the pipe, separates
the scattering region (interior, billiard region, interaction region) 
from the asymptotic region (exterior, pipe).
We ask the average number of times $\langle n \rangle$ a particle
bounces before escaping, or, what is the average dwell time
$\langle \tau \rangle$ of a particle inside the billiard.
As already mentioned, the appropriate measure for averaging
gives equal weights to all points on $\Sigma$ having the
same energy $E$.

In what follows we show that the answers to these questions
are given by very simple ratios between phase space volumes.
Then we argue that our results are also applicable to more general
geometries.

\subsection{Birkhoff maps}

\begin{figure}
\includegraphics[width=8cm]{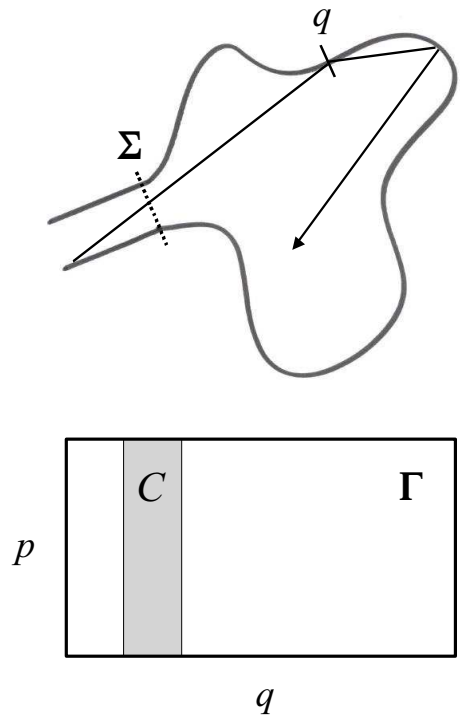}
\caption{%
(top) Billiard with attached pipe.
The Poincar\'e section $\Sigma$ (dashed line) defines an auxiliary
closed billiard.
(bottom) Boundary phase space  $\Gamma$ of the scattering billiard.
The phase space coordinates are $q$, the position along the
boundary, and $p$, its conjugate momentum.
The shaded rectangle $C$ corresponds to the closure $\Sigma$.}
\label{fig1}
\end{figure}

Let us consider the Birkhoff section taken along the billiard
walls, see Fig.~\ref{fig1}.%
The coordinates $(q,p)$, where $q$ is the particle position on
the billiard boundary and $p$ is its conjugate momentum, entirely
characterize the particle phase space $\Gamma$.
The dynamics inside the scattering region is given by the
Birkhoff (or boundary) map $T$ that propagates a particle between
successive bounces, {\it i.e.}, from a phase space point
$(q,p) \in \Gamma$ to the one where the next bounce takes place.

Now we ``close" the billiard by adding a straight segment normal
to the pipe axis \cite{Ozorio00}.
The Poincar\'e section associated to this
segment, $\Sigma$, closes the Birkhoff section $\Gamma$.
Thus, the scattering process can be identified with the
first-recurrence map to $\Sigma$, now considered as a part of
$\Gamma$; the dwell time becomes the first-return time to $\Sigma$.
Figure~\ref{fig1} shows the boundary phase space $\Gamma$,
namely,
a rectangle of length equal to the perimeter of the closed
billiard and height $2p_{\rm max}$, with $p_{\rm max}^2=2mE$,
where $m$ is the particle mass and $E$ its energy. The shaded 
vertical strip corresponds to the closure $\Sigma$, and is 
denoted by $C$.
The inclusion of more pipes to the billiard is accounted
for by adding the corresponding (disjoint) vertical stripes.
This construction can be easily extended to higher dimensions.

We recall that $n$ is the return time measured in units of bounces
against the billiard walls. Its average is 
\be
\label{eq:tclass1}
\langle n \rangle = \frac{\sum_{n=1}^\infty n \, \mu(C_n)}{\mu(C)} \; ,
\ee
where $C_n \subset C$ is the subset of initial conditions that first
return to $C$ after $n$ iterations of the boundary map $T$ and
$\mu$ refers to the volume measure in $\Gamma$.
%
%
Using measure-preservation arguments, it is not difficult to show that
$\sum_{n=1}^\infty n \, \mu(C_n) = 
\mu \left( \bigcup_{n=1}^\infty T^n C \right)$
\cite{Cornfeld82}.
Hence, Eq.~(\ref{eq:tclass1}) becomes
\be
\label{eq:tclass2}
\langle n \rangle = 
\frac{ \mu \left( \bigcup_{n=1}^\infty T^n C \right)}{\mu(C)} \;,
\ee
which now expresses the time delay as a quotient of two measures:
The denominator is the measure of the closure; the numerator represents 
the measure of the inner phase space that
is explored by the ensemble of scattering trajectories.
For an ergodic dynamics the set $\bigcup_{n=1}^\infty T^n C$ clearly
coincides with the full phase space $\Gamma$.
Remarkably, even nonergodic Birkhoff's maps very often satisfy
the {\it weak ergodicity} condition
\be
\label{eq:erg}
\bigcup_{n=1}^\infty T^n C \equiv \Gamma^\prime = \Gamma \;.
\ee
For instance, it is simple to verify that the circle billiard
\cite{Blumel92,Ozorio00}, an archetype of integrable dynamics,
satisfies Eq.~(\ref{eq:erg}) for any straight closure $\Sigma$.
%

As a result, for weakly ergodic billiards
we find that $\langle n \rangle = { \mu(\Gamma)}/{\mu(C)}$.
%
The weak-ergodicity condition is not verified by systems containing
stable islands that cannot be reached from the outside, as for instance,
the cosine-shaped billiard \cite{cosine}.
In these cases we replace $\Gamma$ by $\Gamma^\prime$, the phase space
that is effectively explored by the scattering orbits to write
\be
\label{eq:tclass}
\langle n \rangle = \frac{ \mu(\Gamma^\prime)}{\mu(C)} \; .
\ee
The expression above shows that for both ergodic and non-ergodic systems
$\langle n \rangle$ is finite. Thus, the probability of first returning
after $n$ iterations,
\be
P(n)= \frac{\mu(C_n)}{\mu(C)} \; ,
\ee
must decay faster than $1/n^2$.
In some cases numerical simulations may suggest a divergent average
return time. Note, however, that the true asymptotic decay may settle
only after {\em very} long times \cite{Chirikov99}.
%
%

\subsection{Continuous time}
\label{secB}


The real, continuous, time-delay problem is addressed in analogy to the
simple one presented above.
To make a link between continuous dynamics and maps we invoke the 
stroboscopic map $T_{\Delta t}$, i.e., a discretization of the continuous 
evolution into time steps of length $\Delta t$. The continuum limit
is obtained by making $\Delta t \to 0$. 
The map $T_{\Delta t}$ acts on the full phase space of the scattering
system, namely, a four-dimensional space for a planar billiard. 

In order to adapt Eq.~(\ref{eq:tclass}) to the present context, we note  
that the set of incoming states $C$ has zero measure when thought of as a 
subset of the full phase space: It has to be substituted by a properly
defined set having finite measure, which we call $\overline{C}$. 
The simplest way of choosing $\overline{C}$ is by letting $C$ acquire 
two extra dimensions: in the direction normal to the energy surface, 
and in the direction parallel to the phase space flow. 
The corresponding additional canonical coordinates are the energy 
$E$ and the time $t$ measured along the trajectories starting at the
section $\Sigma$. 
The variables $E$ and $t$, together with the coordinates on $\Sigma$,  
form a local canonical set. 
When $C$ grows in ``thicknesses" by $\Delta E$ and $\Delta t$,
we have the simple relation between measures 
\be
\label{eq:defCbar}
\mu(\overline{C})=\mu(C) \, \Delta E \, \Delta t \; .
\ee
The dwell time is then given by \cite{Balakrishnan00}
\be
\label{eq:tclass-cont}
\langle \tau \rangle  = \lim_{\Delta t \to 0} \Delta t \,
\frac{ \mu(\overline{\Gamma^\prime})}{\mu(\overline{C})} \;,
\ee
where 
\be
\overline{\Gamma^\prime}
=\bigcup_{n=1}^\infty (T_{\Delta t})^n \overline{C} \;.
\ee
The quantity $\overline{\Gamma^\prime}$ represents the inner phase space
for the continuous dynamics that can be accessed from outside.
It has an energy thickness $\Delta E$.

By construction, the set $\overline{C}$ has the important 
property that all its points enter the 
scattering region after one time step $\Delta t$, namely, 
\be
\label{eq:emptyintersection}
\mu(\overline{C} \cap T_{\Delta t}\overline{C})=0 \; .
\ee
This avoids the problem of having to subtract spurious 
contributions to the dwell time arising from non-scattering orbits 
\cite{Balakrishnan00}.

Let us now express Eq.~(\ref{eq:tclass-cont}) in terms of more
appealing quantities.
Define $\Omega$ to be the phase space volume contained by the energy 
shell $E$ within the scattering region 
(as before, primes will indicate ``accessible from outside").
Then
\be
\mu(\overline{\Gamma^\prime}) = 
\frac{\partial  
\Omega^\prime}{\partial E} \Delta E \; .
\ee
We recall that $\mu(C)$ is the phase space volume contained 
by the energy shell $E$ {\em within the section} $\Sigma$. 
We now switch to a more standard notation and, from now on, 
we call it $\Omega_\Sigma$.
Gathering everything and substituting into Eq.~(\ref{eq:tclass-cont}) 
we arrive at
\be
\label{eq:levinson}
\langle \tau \rangle
=\frac{1}{\Omega_\Sigma}\frac{\partial \Omega^\prime}{\partial E}  \; .
\ee
This remarkable formula is exact and holds irrespective of the 
dynamics being chaotic, regular or mixed. 
After making the proper identifications, Eq.~(\ref{eq:levinson}) 
can also be applied to billiard systems in three (or higher) 
dimensions. 

For the sake of illustration, let us, for instance, use 
Eq.~(\ref{eq:levinson}) to calculate the mean time between 
collisions for a closed billiard. 
In this case, we have to choose $\Sigma$ as the phase space 
corresponding to the full boundary $\cal{L}$. 
The weak ergodicity condition is obviously satisfied and the 
average bounce time reads
$\tau_{\rm b}=(\partial \Omega/\partial E)/\Omega_{\cal{L}}$
for any billiard, truly ergodic or not. 
For the two-dimensional case, 
$\Omega_{\cal{L}} = 2 m v   {\cal L}$ and 
$\Omega = \pi p^2 {\cal A} = 2 \pi m E {\cal A}$
($v$ = velocity, ${\cal L}$ = perimeter, ${\cal A}$ = area).
Then we arrive at the well-known result
$\tau_{\rm b}=  \pi {\cal A}/v {\cal L}$,
as heuristically shown in Ref.~\cite{Jarzynski93} 
and proven in Ref.~\cite{Chernov97}.

\subsection{Smooth systems}

The extension of our findings to smooth systems is immediate.
The boundary that so far defined the system billiard-plus-pipe 
now is thought as the level curve of a smooth potential.
The motion in the waveguide is free in the longitudinal direction
($\hat x$). 
In the directions transverse to the waveguide ($\hat y$) the dynamics
is governed by a smooth Hamiltonian $H_\perp(y,p_y)$. 

The analysis of Sec.~\ref{secB} applies equally well to this case
Thus, the formula for the dwell time is also Eq.~(\ref{eq:levinson}), 
with the following definitions. 
$\Omega_\Sigma$ is the measure of the phase space in the Poincar\'e 
section lying inside the energy shell $H_\perp = E$,
\begin{equation}
\Omega_\Sigma = \int_{H_\perp \le E} dy \, dp_y \; .
\end{equation}
$\Omega$ is the volume of the inner phase space with energy less
than $E$. Assuming that the scatterer lies in the region $x>0$, we
have
\be
\Omega = 
\int_{H \le E \;\; {\rm and} \;\; x>0 } dx \, dy \, dp_x \, dp_y \; .
\ee

The case of a particle scattered off a smooth potential in three dimensions 
can be accounted for by enclosing the scatterer with a large enough spherical 
shell (the Poincar\'e section $\Sigma$). Then one defines the delay time 
as the (average) return time to $\Sigma$ minus the return time when there
is no potential. Both return times are special instances of 
Eq.~(\ref{eq:levinson}), the free-flight time being just the average 
bounce time of a spherical billiard.

\section{Average Wigner time-delay}
\label{sec:quantum-time-delay}

The Wigner time delay $\tau_W(E)$, given by 
Eq.\ (\ref{eq:Wigner-time-delay}), fluctuates as a 
function of the energy.
Large time delays are due to resonant scattering, whereas off-resonance 
scattering corresponds to direct processes that spend short times in the 
interaction region. 
This picture becomes particularly clear in the regime of isolated 
resonances: 
Long time delays occur at narrow energy windows around each resonance, 
in the remaining energy interval scattering processes are fast (direct). 
The important energy scale that emerges from this picture is the mean 
resonance spacing. 
When the resonances are overlapping, the separation of time scales is less 
clear, and fluctuations are much smaller \cite{Lewenkopf91}.

By averaging the Wigner time delay over an energy window $\Delta E$ 
containing many resonances, fast and slow processes concur to
give a very simple expression:
\be
\langle \tau_W(E) \rangle = \frac{h}{N} \rho(E) \;,
\label{eq:avWigner}
\ee
where $\rho(E)$ is the mean resonance density (due to the scattering
region).
Hence, $\langle \tau_W \rangle$ basically just counts the number 
of resonances within the energy interval $\Delta E$.
Equation~(\ref{eq:avWigner}) can be derived in various ways, for instance,
using the ${\sf S}$--matrix pole structure \cite{Lewenkopf91}.

In order to relate $\langle \tau_W \rangle$ with the classical results, 
we take the semiclassical limit of Eq.~(\ref{eq:avWigner}).
We first use the Weyl formula to express the mean resonance density 
$\rho(E)$.
For that purpose, we consider the corresponding closed system (scattering 
region closed by $\Sigma$), to write 
\be
\widetilde{\rho}(E) = \frac{1}{h^d}\frac{\partial \Omega}{\partial E} \; ,
\ee
where $d$ is the dimension of the system. The wide tilde is used to indicate 
that the semiclassical limit was taken.
By the same token, the number of states in the pipes is given by 
\be
\widetilde{N}=\frac{\Omega_\Sigma}{h^{d-1}}   \; .
\ee
We then arrive at
\be
\label{eq:tauWsc}
\langle \widetilde{\tau}_W \rangle 
=\frac{1}{\Omega_\Sigma}\frac{\partial \Omega}{\partial E}  \; .
\ee
Remarkably, as in the classical case, the average Wigner time delay 
is a purely geometric quantity, and does not capture dynamical features.

\section{Conclusions}
\label{sec:conclusions}

The most striking result of our semiclassical analysis in that 
the Wigner time delay of Eq.~(\ref{eq:tauWsc}) is not 
in correspondence with the classical dwell time of 
Eq.~(\ref{eq:levinson}). 
The correspondence holds only in the case of weak ergodicity, 
where the phase space volume $\Omega^\prime$ equals $\Omega$.
Both quantities are different in the more general situation of a 
mixed phase space.

This result can be interpreted as follows:
In general, mixed systems have phase space domains in the 
interaction region which are not classically accessible 
from the outside. 
These regions, if larger than $h^d$ will support quantum states.
Such states correspond to resonances, that can be very thin, 
depending on the height of the dynamical tunneling barriers.
As we showed, they contribute to $\langle \tau_W \rangle $ with the 
same weight as other quantum states predominantly localized in 
classically accessible regions.

In a broader picture, we speculate that the lack of classical-quantum
correspondence for the dwell time is another manifestation of the 
non commutativity between the long time limit ($t \rightarrow \infty$) 
and the semiclassical limit ($\hbar \rightarrow 0$). 
Tunneling into (or out from) localized states at islands of the mixed 
phase space takes a very large time scale to occur, and is absent in 
the classical limit of $\hbar = 0$.

We conclude by stressing that our results are rigorous, and do not 
depend on the here presented interpretations.

\begin{acknowledgments}
We thank A. M. Ozorio de Almeida for many interesting comments.
Partial financial support from CNPq and PRONEX is gratefully acknowledged. 
\end{acknowledgments}


\end{document}